\documentclass[12pt]{article}

\usepackage{newtxtext,newtxmath}
\usepackage{graphicx}

\usepackage[letterpaper,margin=1in]{geometry}

\renewenvironment{abstract}
	{\quotation}
	{\endquotation}

\date{}

\makeatletter
\renewcommand{\fnum@figure}{\textbf{Figure \thefigure}}
\renewcommand{\fnum@table}{\textbf{Table \thetable}}
\makeatother

\usepackage{scicite}

\usepackage{url}

\def\scititle{
	West Antarctic Meltwater can Prevent an AMOC Collapse
}
\title{\bfseries \boldmath \scititle}

\author{
	Sacha Sinet$^{1,2\ast}$,
	Anna. S. von der Heydt$^{1,2}$,
	Henk A. Dijkstra$^{1,2}$\and
	\small$^{1}$Department of Physics, Institute for Marine and Atmospheric research, \\ \small Utrecht University,  Utrecht, 3584 CC, the Netherlands.\and
	\small$^{2}$Centre for Complex Systems Studies, Utrecht Univeristy, Utrecht, 3584 CC, the Netherlands.\and
	\small$^\ast$Corresponding author. Email: s.a.m.sinet@uu.nl 
}

\begin{document} 

% Insert the title and author list
\maketitle

% Abstract, in bold
% There are strict length limits, and not all formats have abstracts.
% Consult the journal instructions to authors for details.
% Do not cite any references in the abstract.
\begin{abstract} \bfseries \boldmath
% Start with one or two sentences of background
\noindent The Atlantic Meridional Overturning Circulation (AMOC) and polar ice sheets are coupled tipping elements, allowing for potential cascading tipping events in which tipping is facilitated by their mutual interactions. However, while an AMOC destabilization driven by Greenland Ice Sheet (GIS) meltwater release is well documented, the consequences of a West Antarctic Ice Sheet (WAIS) tipping on the AMOC remain unclear. In the Earth System Model of Intermediate Complexity CLIMBER-X, we perform experiments where meltwater fluxes representing plausible tipping trajectories of the GIS and WAIS are applied. We find that WAIS meltwater input can increase the AMOC resilience to GIS meltwater. In particular, we show that this stabilizing effect can cause the AMOC recovery and, for the first time in a comprehensive model, totally prevent an AMOC collapse. Moreover, we find this stabilzation to occur for ice sheet tipping trajectories that are relevant under high future greenhouse gas emission scenarios.
\end{abstract}

% The first paragraph of any Science paper does NOT have a heading
% Nor is it indented

\section*{Introduction}
\noindent The evergrowing threat of climate change may be exacerbated by the existence of critical thresholds or tipping points within the Earth system \cite{lenton_tipping_2008}. Once crossed, abrupt and irreversible shifts may occur in parts of the climate system known as tipping elements, of which many are crucial in maintaining the functioning and stability of the present-day climate, ecosystems and societies. 

One of these tipping elements is the Atlantic Meridional Overturning Circulation (AMOC), a large system of ocean currents redistributing heat, salt and nutrients between the equator and the poles. Widely sustained by the sinking of dense waters in the northern part of the Atlantic Ocean, its tipping dynamics are mainly understood in terms of the salt-advection feedback \cite{weijer_stability_2019}. An initial freshening of the North Atlantic results in less sinking of dense waters in this region. This leads to a weaker AMOC transporting less salt to the North Atlantic, hence amplifying the initial perturbation. Consequently, the stability of the AMOC is tightly connected to the future development of polar ice sheets, for which the mass loss (and therefore meltwater release into the ocean) is likely to increase throughout this century \cite{foxkemper2021}. In particular, the Greenland Ice Sheet (GIS) and West Antarctic Ice Sheet (WAIS) are also classified as tipping elements, and are at risk of undergoing an irreversible collapse already at the current level of warming \cite{armstrong_mckay_exceeding_2022}. 

On one hand, it sets the ground for potentially dangerous domino effects referred to as cascading tipping events \cite{dekker_cascading_2018,klose_what_2021,wunderling_climate_2024}. In such an occurrence, the tipping of the whole system may occur while global warming only pushes one component beyond its critical threshold. A typical worst case scenario goes as follows. In a warming world, the GIS may be destabilized via different processes such as the melt-elevation feedback \cite{weertman_stability_1961, levermann_simple_2016}, resulting in substantial meltwater release into the North Atlantic. This freshwater perturbation may then be amplified by the salt-advection feedback and result in a weakening or tipping of the AMOC \cite{jackson_understanding_2023}. Finally, the subsequent warming of the Southern Hemisphere \cite{orihuela-pinto_interbasin_2022,jackson_global_2015} may trigger postitive feedbacks such as the marine ice sheet instability, destabilizing the WAIS \cite{joughin_stability_2011}. On the other hand, this intricate system includes interactions which may result in the tipping of one component to be beneficial for another. In the aforementioned example, the AMOC weakening has been shown to imply a substantial cooling over the North Pole \cite{jackson_global_2015,van_westen_physics-based_2024,orihuela-pinto_interbasin_2022}, which could inhibit a GIS collapse. Hence, a thorough understanding of the individual dynamics and interactions driving this coupled system of tipping elements is crucial for exploring the range of possible consequences of climate change.

A key uncertainty in how cascading tipping may or not unfold in this system lies is how a WAIS tipping could impact the AMOC stability. Indeed, most model studies find that the AMOC is only moderately affected by freshwater release into the Southern Ocean \cite{swingedouw_impact_2009,seidov_is_2005,stouffer_climate_2007}, while some studies have pointed out that such an event may delay a weakening of the AMOC \cite{sadai_future_2020} or even help it to recover from a total collapse \cite{weaver_meltwater_2003}. More recently, two different studies found that a collapse of the WAIS may result in AMOC tipping to be totally avoided \cite{sinet_amoc_2023,sinet_amoc_2023-1}. However, the AMOC response to meltwater release in the Southern Hemisphere was highly simplified, calling for investigation in models of higher complexity. 

In this study, we show the different AMOC responses that can occur as a consequence of a collapse of both the GIS and WAIS using the Earth System Model of Intermediate Complexity (EMIC) CLIMBER-X v1.0. For this, we perform experiments in which the freshwater flux is parametrized to represent ice sheet tipping trajectories occurring on a realistic timescale. In particular, we show that an AMOC tipping prevented by WAIS meltwater as found in \cite{sinet_amoc_2023,sinet_amoc_2023-1} also exists in CLIMBER-X, and provide a mechanistic explanation of this phenomenon.

\section*{Results}
Our experiments are designed to investigate the AMOC response to ice sheet tipping events occurring on a plausible timescale. However, predicting long-term (tipping) trajectories for polar ice sheets is a major challenge, which owes from uncertainties originating from ice sheet modelling, but also from future greenhouse gas emissions and the associated climate sensitivity \cite{foxkemper2021}. In general, the rate of melting increases with the level of warming, such that ice sheet tipping events occurring on the fastest timescale are typically associated with high to very high emission scenarios. It is the case of the Representative Concentration Pathway (RCP) 8.5, for which modelling studies suggest that a collapse of the GIS and WAIS could occur as fast as in 1000 \cite{aschwanden_contribution_2019} and 500 \cite{deconto_contribution_2016,chambers_mass_2022} years, respectively. These are the lower bounds proposed by recent literature, in which plausible tipping durations for the GIS were given between 1000 and 15000~years, while tipping durations for the WAIS were given between 500 and 13000~years \cite{armstrong_mckay_exceeding_2022}. Within these relevant ranges, we design hosing experiments using the EMIC CLIMBER-X v1.0 (see Methods), in which the meltwater forcing induced by ice sheet tipping is captured similarly to what was done in previous literature \cite{sinet_amoc_2023-1}. Namely, these tipping events are parametrized only by their duration and onset delay, and inserted without compensation around the GIS and WAIS to conceptually capture a full collapse of both polar ice sheets (see Methods). 

\subsection*{Separate Greenland and West Antarctica meltwater impacts}
\begin{figure}
    \centering
    \includegraphics[width = 1.0\textwidth]{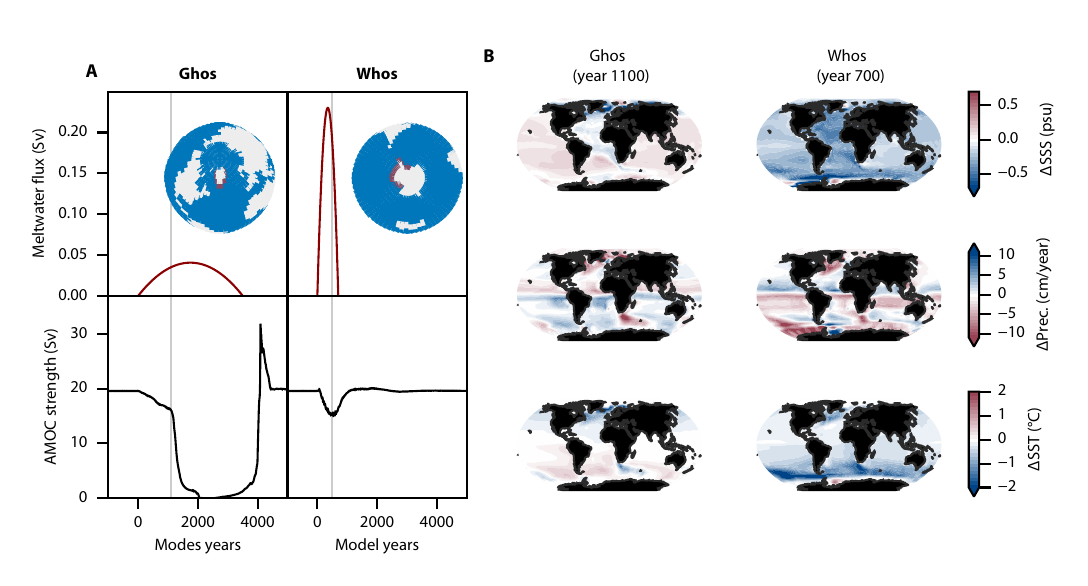}
    
    \caption{\textbf{AMOC  and climate response to separate GIS and WAIS meltwater fluxes.} (\textbf{A}) Meltwater flux (red) and AMOC trajectory (black) in the Ghos (left column) and Whos (right column) experiments, respectively. (\textbf{B}) Sea surface salinity anomaly (above), precipitation anomaly (middle) and sea surface temperature anomaly (below). These are yearly averages taken at year~1100 in the Ghos experiment (left column) and at year~500 in the Whos experiment (right column), indicated by grey vertical lines in panel A.}
    \label{fig:separate}
\end{figure}

\noindent Understanding the AMOC response to the combined meltwater fluxes from the GIS and WAIS requires to identify their isolated impacts. First, applying only a meltwater flux from the GIS, we find that an AMOC tipping occurs for any GIS collapse trajectory lasting between 1000 and 3900~years. In Figure \ref{fig:separate}.A (left), the experiment in which an AMOC tipping is induced by a GIS meltwater forcing lasting 3500~years (denoted Ghos, see table \ref{tab:exp}), which is of the order of the time in which the GIS may become ice free in high to very high emission scenarios \cite{clark_consequences_2016, van_breedam_semi-equilibrated_2020}, is shown. From the initiation of the GIS meltwater forcing (at year 0 by convention), an AMOC weakening is found up to about year 1100, after which the AMOC rapidly collapses to OFF state as the GIS meltwater forcing reaches 0.035 Sverdrup ($1~\text{Sv}= 10^6~\text{m}^3/\text{s}$). Indeed, the meltwater insertion induces a widespread freshening of the surface waters in the Northern Atlantic, as can be seen from the Sea Surface Salinity (SSS) anomaly juste before AMOC tipping in Figure \ref{fig:separate}.B (upper left). This freshening results in a density decrease in the North Atlantic Deep Water (NADW) formation regions, thereby altering deep convection, weakening the AMOC and leading to its collapse as the positive salt-advection feedback takes over. Consistent with previous literature using higher complexity models, the AMOC weakening induces both a southward shift of the Intertropical Convergence Zone (ITCZ) (Figure \ref{fig:separate}.B, middle left), as well as a cooling and warming of the Northern and Southern Hemisphere, respectively (Figure \ref{fig:separate}.B, lower left)\cite{jackson_global_2015,van_westen_physics-based_2024, orihuela-pinto_interbasin_2022}. Second, from about year 3000, the AMOC slowly recovers from the OFF state as the GIS meltwater forcing decreases, stabilizing at present-day level from about year 4500. During this phase, a very weak overturning circulation remains and, as the GIS meltwater flux decreases, initiates a positive salt-advection feedback. Namely, as salty equatorial surface waters are slowly advected northward, the progressive density increase of high latitude North Atlantic surface waters eventually results in a reactivation of deep convection. We note that the AMOC recovery at the forcing termination is commonly found in EMICs \cite{rahmstorf_thermohaline_2005}, and that the substantial overshoot of the AMOC initial value by more than 10~Sv during its recovery phase was recently found in the Community Earth System Model (CESM) \cite{van_westen_asymmetry_2023}.

Second, we isolate the effects of a meltwater flux originating from the WAIS. While we find no AMOC tipping for any value of the WAIS tipping duration higher than 500~years, a sustained AMOC weakening is systematically found after a short period of strengthening, consistently with most previous literature using models of similar complexity \cite{an_antarctic_2024, swingedouw_impact_2009, stouffer_climate_2007}. In Figure \ref{fig:separate}.A (right), the experiment in which an AMOC weakening of about 4.5~Sv is induced by a WAIS meltwater forcing lasting 700 years (denoted Whos, see table \ref{tab:exp}) is shown. As for the Ghos experiment, it compares to the time taken by the WAIS to become ice free in high to very high emission scenarios \cite{deconto_contribution_2016, chambers_mass_2022}. At the weakest AMOC state (at about year 500), we find a worldwide surface water freshening especially pronounced in the North Atlantic (Figure \ref{fig:separate}.B, upper right) which is well established by previous literature, and in itself suffices to explain the AMOC weakening \cite{an_antarctic_2024,stouffer_climate_2007,seidov_is_2005, swingedouw_impact_2009}. Moreover, the AMOC strength decrease is mostly proportional to the applied meltwater flux, although with a delay of a few decades, which is about the time needed for the Southern Ocean surface freshening to propagate into the North Atlantic ocean \cite{an_antarctic_2024,swingedouw_impact_2009}. Finally, consistently with previous literature, we find a northward shift of the ITCZ (Figure \ref{fig:separate}.B, middle right) \cite{bakker_response_2018, an_antarctic_2024} as well as a pronounced cooling of the Southern Hemisphere (Figure \ref{fig:separate}.B, lower right) \cite{an_antarctic_2024,sadai_future_2020,li_global_2023}.

\subsection*{AMOC resilience modified by West Antarctic Meltwater}
\begin{figure}[h!]
    \centering
    \includegraphics[width=1.0\textwidth]{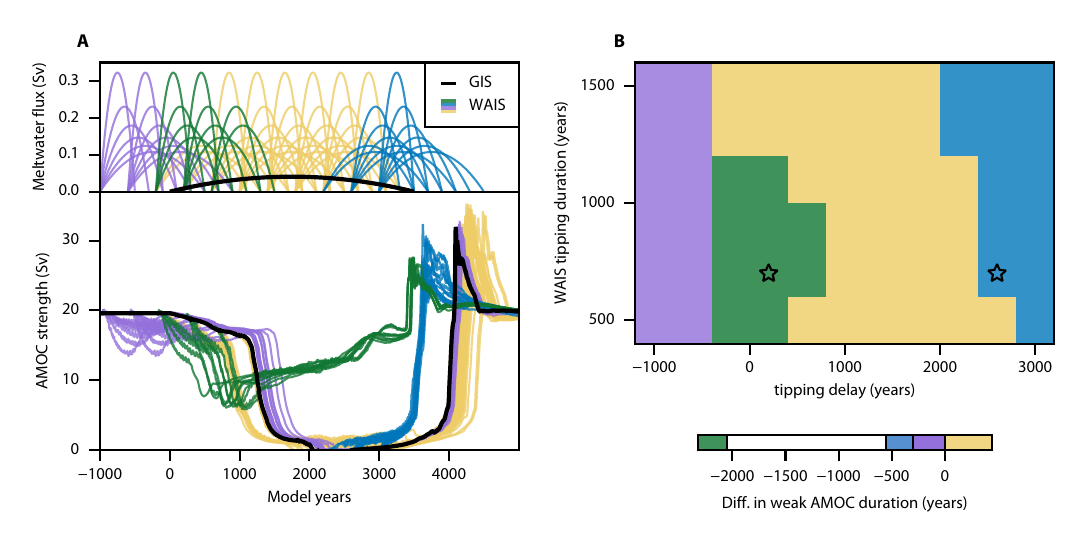}
    
    \caption{\textbf{AMOC response to combined GIS and WAIS meltwater fluxes.} (\textbf{A}) Trajectories of the meltwater fluxes (above) and AMOC (below), fixing the GIS tipping duration to 3500~years. The black trajectory is forced only by the GIS meltwater flux (Ghos experiment, see table \ref{tab:exp}), while these in colours include meltwater from both the GIS and WAIS. Colours represent different values of the difference in weak AMOC duration $\Delta T_\text{weak}$. (\textbf{B}) Difference in weak AMOC duration $\Delta T_\text{weak}$ for trajectories displayed in A, represented in the WAIS tipping durations and tipping delays parameter plane.}
    \label{fig:res}
\end{figure}

To understand the combined impacts of applying both GIS and WAIS meltwater fluxes on the AMOC stability, we will consider the experiment in which only the GIS meltwater forcing lasting 3500~years induced an AMOC collapse (Ghos, see table \ref{tab:exp}) as a baseline experiment. Then, using the same GIS meltwater forcing, we add meltwater fluxes originating from the WAIS using different values of both the WAIS collapse duration and the delay between ice sheet tipping events. To quantify the (change in) AMOC resilience, we will use the duration of the weak AMOC state denoted $T_\text{weak}$, here chosen as the time during which the AMOC strength is below 10~Sv (to be thought of a characteristic return time, a classical measure of resilience in dynamical systems \cite{Krakovska_Kuehn_Longo_2024}). In particular, we quantify the change of AMOC resilience induced by the WAIS meltwater flux via the following metric
\begin{equation}
	\Delta T_\text{weak} = T_\text{weak} -  T_\text{weak}^0,
\end{equation}
where $T^0_\text{weak}$ is the weak AMOC duration computed in the Ghos experiment. In other words, $\Delta T_\text{weak}$ is the amount of years by which the weak AMOC state duration has been varied, with negative values indicating a more resilient AMOC. In Figure \ref{fig:res}.A, the AMOC response to the GIS meltwater flux lasting 3500~years (Ghos experiment) is shown in black, while coloured lines are trajectories that also include meltwater flux for the WAIS. These colours stand for different value of $\Delta T_\text{weak}$, which are also presented in the parameter space defining the different WAIS trajectories, namely the WAIS tipping duration and delay between ice sheet tipping events (Figure \ref{fig:res}.B). We find that the WAIS meltwater forcing drastically impacts the AMOC resilience to the GIS meltwater flux.

~\\
 On one hand, a WAIS meltwater flux initiated roughly after year 1000, i.e. applied on the AMOC OFF state, may both delay or accelerate the AMOC recovery process. In cases of an accelerated recovery (blue), the AMOC gains about 2~Sv as soon as the WAIS meltwater insertion starts, at a moment when the GIS meltwater flux decreases fast enough for this weak AMOC to be sustained. This provides an earlier activation of the positive salt advection feedback required for the AMOC recovery, which occurs up to about 500~years earlier. This recovery induced by WAIS meltwater insertion has already been found and explained in previous literature \cite{weaver_meltwater_2003}. Namely, the freshening and hence lightening of the Antarctic Intermediate Waters (AAIW) reestablishes a density contrast between the AAIW and NADW which is favourable for the overturning circulation, resulting in the AMOC reactivation. Consistently, the lightening of the AAIW is also clearly found in our results (figure \ref{fig:reco}), as can be seen in the GWhosR experiment (blue star in Figure \ref{fig:res}.B, see table \ref{tab:exp}). In the case of a delayed recovery (yellow), we find the same initial kick to the AMOC for about the duration of the WAIS meltwater forcing. However, the GIS meltwater flux is still too strong, such that it is soon attracted back to the OFF state, delaying the AMOC recovery process. In fact, these two regimes appear to be well discriminated by the moment at which the maximum of the West Antarctic meltwater flux occurs, as we find an accelerated recovery to occur in cases where the peak WAIS meltwater flux occurs approximately after year 3000.

 On the other hand, we find that a WAIS meltwater flux initiated before year 1000, i.e. on the AMOC ON state, systematically results in a faster AMOC weakening compared to the Ghos experiment. This is expected from the AMOC weakening implied by both the GIS and WAIS meltwater fluxes as found in Figure \ref{fig:separate}.A, and can result in an accelerated AMOC tipping (yellow). Yet, we also find that some WAIS tipping trajectories initiated early (before or at year $-400$) result in a delayed AMOC collapse (purple). Remarkably, we find that some WAIS tipping trajectories result in the AMOC collapse to be totally avoided (green). In these stabilization cases, the AMOC engages in a steady recovery as soon as it is decreased by a maximum 60 percent, resulting in an AMOC weak state duration decreased by as much as 2100~years. Qualitatively, the parameter region in which this stabilization is found appears consistent with previous conceptual research \cite{sinet_amoc_2023-1}. Indeed, we find that stabilization occurs for rather short and strong WAIS meltwater fluxes, and in cases where the peak of the WAIS meltwater flux occurs about 1000~years before the peak of the GIS meltwater flux.x We note that the disappearance of this stabilization region can be expected from a faster GIS tipping trajectory yielding a stronger meltwater flux. To show this, we perform a similar set of experiments for a GIS meltwater flux lasting 1000 rather than 3500~years, to be thought of as a worst-case scenario \cite{aschwanden_contribution_2019,armstrong_mckay_exceeding_2022}. In this case, while the earlier recovery and tipping regimes are still found, WAIS meltwater fluxes are only able to marginally delay the tipping, while the stabilization regime totally disappears within the tested parameter range (figure \ref{fig:ressupp}). Finally, considering the results using a GIS tipping duration of 3500~years (Figure \ref{fig:res}), the AMOC stabilization occurs for ice sheet tipping trajectories that are relevant considering high to very high future emission patways. Indeed, an upper bound for the tipping point of both the GIS and WAIS at 3~$^\circ$C above pre-industrial levels was proposed by recent literature \cite{armstrong_mckay_exceeding_2022}. This suggests that the tipping of both ice sheets is likely to occur within the next century in the case of the Shared Socioeconomic Pathways (SSP) 5-7.0 or higher emission scenarios. This would translate into a delay between the initiation of both tipping events of maximum a century which, in our experiments, is the most favourable situation for the stabilzation to occur, requiring a rather fast WAIS tipping lasting no more than 1100~years.

\subsection*{Mechanisms of AMOC stabilization}
To understand how the stabilization process occurs, we proceed to a comparison between two experiments. On one hand, we consider the Ghos experiment (see table \ref{tab:exp}) in which only a GIS meltwater flux lasting 3500~years yields an AMOC collapse. On the other hand, we consider the GWhosS experiment  (green star in Figure \ref{fig:res}.B, see table \ref{tab:exp}) in which the same GIS meltwater flux is applied, while a WAIS meltwater flux lasting 700~years and initiated 200~years after the GIS meltwater flux yields an AMOC stabilization. For the purpose of our analysis, we compute the average density of two boxes in the Atlantic Ocean, representing the Northern (45-60~$^\circ$N) and Southern (35-20~$^\circ$S) Atlantic, denoted the NAtl and SAtl box, respectively, and compute the full freshwater balance of the NAtl box (figure \ref{fig:FWsupp}, see Methods). In this context, the stabilization process can be understood in the following three stages.

\begin{figure}[t!]
    \centering
    \includegraphics[width=1.0\textwidth]{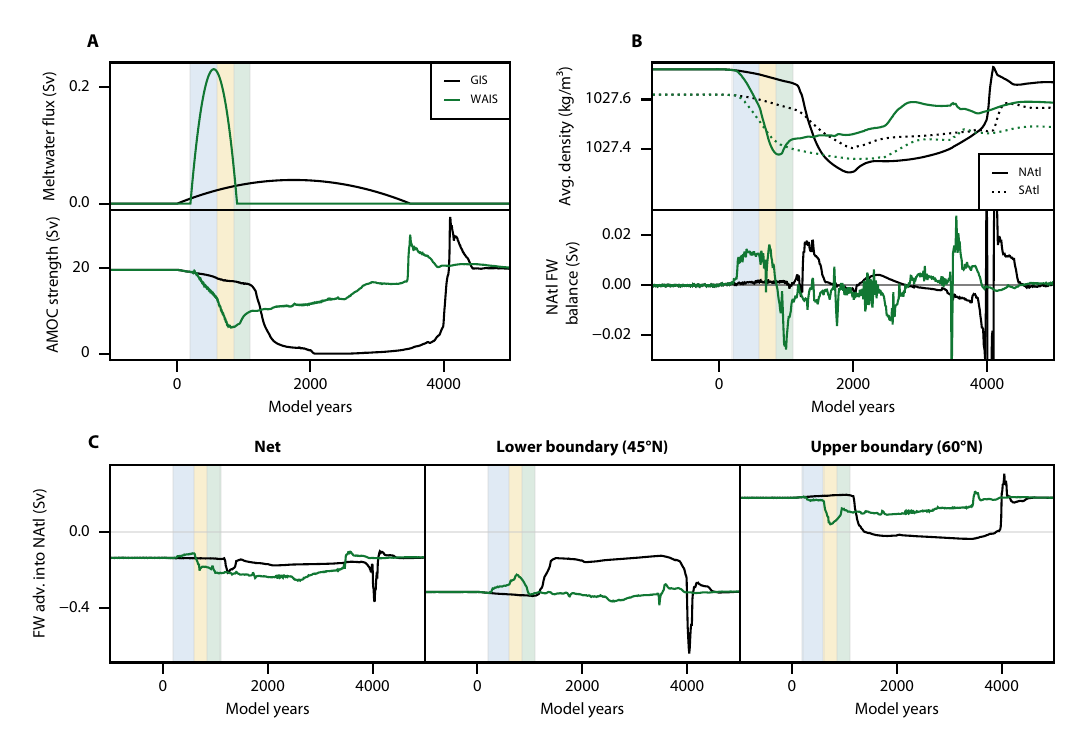}
    
    \caption{\textbf{AMOC stabilization driven by West Antarctic meltwater.} (\textbf{A}) Trajectory of the meltwater fluxes and AMOC strength in the Ghos (black) and GWhosS (green) experiments, the latter including both the GIS and WAIS meltwater fluxes. (\textbf{B}) Average density of the NAtl and SAtl boxes (above) and freshwater balance of the NAtl box (below), displayed for both the Ghos (black) and GWhosS (green) experiments. (\textbf{C}) Net freshwater advection into the NAtl box (left), along with the advection into the Natl box at its lower (middle) and upper (right) boundaries. These are displayed for both Ghos (black) and GWhosS (green) experiments.}
    \label{fig:weak}
\end{figure}

\begin{figure}[t!]
    \centering
    \includegraphics[width=1.0\textwidth]{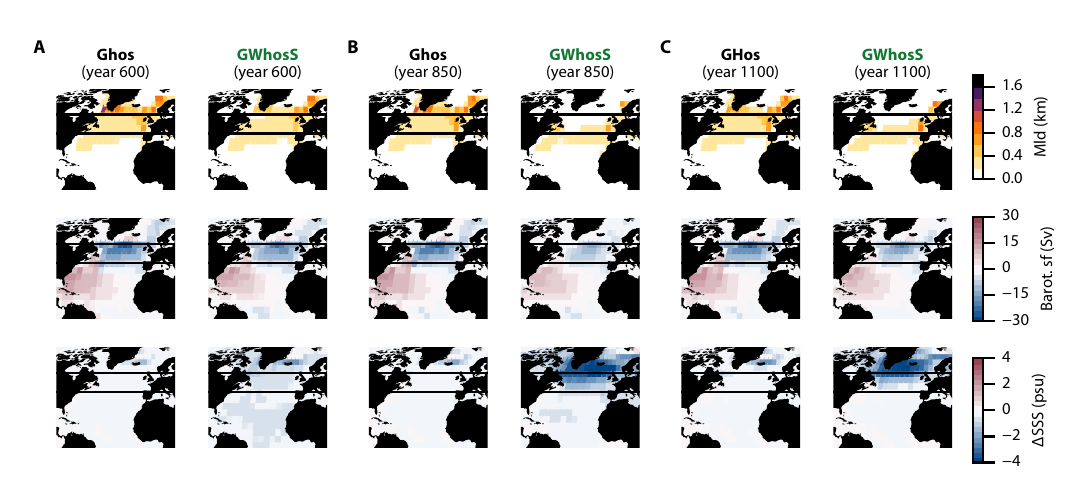}
    
    \caption{\textbf{AMOC diagnostics at the end of each stabilization phases.} (\textbf{A}) Diagnostics of the mixed layer depth (Mld), barotropic stream-function and SSS anomaly at the end of the first stabilization phase (year 600) for the Ghos (left) and GWhosS (right) experiments. The upper and lower bounds of the NAtl box are represented as horizontal black lines. (\textbf{B}) Same as A but at the end of the second stabilization phase (year 850). (\textbf{C}) Same as A but at the end of the third stabilization phase (year 1100).}
    \label{fig:weaksecond}
\end{figure}

In a first stage (between years 200 and 600, light blue band in Figure \ref{fig:weak}), we find an accelerated weakening of the AMOC in the GWhosS experiment (Figure \ref{fig:weak}.A, below). As previously explained, this mostly originates from the meltwater discharge around the WAIS being advected into the Atlantic Ocean, and in particular into the North Atlantic. This is evident from the increase of advective freshwater import at the lower boundary of the NAtl box (Figure \ref{fig:weak}.C, middle), which can be further identified as being advected northward by the overturning circulation itself (figure \ref{fig:FWsupp}, lower middle). At the end of the first stage (Figure \ref{fig:weaksecond}.A), the increased surface freshening in the North Atlantic compared to the Ghos experiment is clear, along with a weakening of both deep convection and the gyre circulation, as can be seen from the mixed layer depth and barotropic stream function, respectively. We note that the weakening of the gyre circulation as the AMOC strength decreases is also found in the higher hierarchy of models \cite{mimi_atlantic_2024}.

This leads to a second stage (between years 600 and 800, light yellow band in Figure \ref{fig:weak}), in which the GWhosS experiment yields a sharp decline of the AMOC strength as convection in high latitudes collapses. This occurs as the gyre circulation, which normally transports the freshwater from the north Atlantic to lower latitudes, is now too weak for advecting the increasing freshwater buildup forming around the GIS southward (Figure \ref{fig:weak}.C, right and figure \ref{fig:FWsupp}.C, upper and lower right). This leads to accumulation of freshwater around Greenland, thereby decreasing the surface water density there, and stopping deep convection as well as weakening the gyre circulation further. At the end of the second phase, the collapse of deep convection in the sub-polar region, weakening of the gyres as well as freshwater buildup around the GIS, are established (Figure \ref{fig:weaksecond}.B).  

In a third stage (between years 800 and 1100, light green band in Figure \ref{fig:weak}), the AMOC recovers and stabilizes to a weak state. On one hand, we find that this recovery from year 800 is contingent with a sharp density increase in the North Atlantic (Figure \ref{fig:weak}.B, above). This occurs as, at this point, the weak gyre circulation only advects a fraction of the GIS freshwater southward (Figure \ref{fig:weak}.C, right), resulting in most of this freshwater to accumulate at high latitudes, thereby preserving the weak deep convection that still occurs under 55~$^\circ$N. Meanwhile, as the meltwater flux originating from the WAIS weakens, the freshwater export out of the NAtl box at its southern boundary becomes similar to present day (Figure \ref{fig:weak}.C, middle). As seen from the freshwater balance (Figure \ref{fig:weak}.B, below), this results in a net freshwater export out of the NAtl box, meaning an increasing salinity and hence density there (Figure \ref{fig:weak}.B, above). On the other hand, as the fraction of the increasing GIS freshwater flux reaching the NAtl box is advected southward, the density of the SAtl box decreases (Figure \ref{fig:weak}.B, above). In summary, both the increase of density in the North Atlantic and increasing meridional density gradient between the North and South Atlantic contribute to maintain the AMOC in a weaker, but active state which gains strength as the GIS meltwater forcing decreases. Notably, in the same model, this weak AMOC state characterised by convection occurring below 55N (Figure \ref{fig:weaksecond}.C) has already been found in recent literature, where it was shown to be stable at these higher values of the northern meltwater flux \cite{willeit_generalized_2024}. 

\section*{Discussion}
Our results shed new light on the impact that a future GIS and WAIS tipping may have on the AMOC. While the destabilising influence of a GIS meltwater discharge is found consistently with previous literature, we identify qualitatively different regimes that can occur when a WAIS tipping is also represented. In the context of a GIS-induced AMOC tipping, we demonstrate that WAIS meltwater is able to totally prevent the AMOC collapse, an outcome initially found in conceptual studies \cite{sinet_amoc_2023,sinet_amoc_2023-1} and now for the first time in a comprehensive climate model. 

We also provide a mechanistic understanding of this stabilization phenomenon. In a first phase, the northward meltwater export from the WAIS results in a faster AMOC weakening leading, in a second phase wherein the GIS meltwater flux becomes stronger, to an early collapse of the circulation at high latitudes. This sets the AMOC in a weaker but more resilient stable state, in which the GIS meltwater accumulating around the GIS has only limited impact on convection regions active at lower latitudes. This alternative equilibrium that was already found earlier \cite{willeit_generalized_2024} results, in a third phase, to the AMOC recovery as the GIS forcing dereases. Similarly to previous studies \cite{sinet_amoc_2023,sinet_amoc_2023-1}, stabilization occurs for a rather fast collapse of the WAIS lasting up to 1100~years, and in cases where the peak of meltwater flux occurs about thousand years before the peak of GIS meltwater flux.

The wide range of WAIS tipping trajectories explored in this study can be interpreted in light of both future and past climate. In terms of future emission scenarios and in the case of a tipping of both polar ice sheets, plausible tipping points proposed by literature \cite{armstrong_mckay_exceeding_2022} would result in both ice sheets to collapse within the 21st century for high emission scenario SSP5-7.0 to 8.5, or within the next 300~years for milder emission scenarios such as SSP2-4.5 when considering extended SSP scenarios \cite{lee2021}. Hence, any of these scenarios would result in a delay between the initiation of ice sheet tipping at both poles of at most a few centuries, resulting in most likely future scenarios to be either an accelerated or prevented AMOC collapse. Yet, the rather fast GIS tipping lasting 3500~years used in our experiments makes it safer to interpret our results in the context of high to very high emission scenarios. In this case, a fast and early collapse of the WAIS on the (sub)millennial timescale is possible, resulting in the AMOC stabilization phenomenon to be a credible possibility. Meanwhile, scenarios of a WAIS collapse acting on a tipped AMOC (i.e. for tipping delays greater than 2000~years) are relevant in the context of paleo-climate. Indeed, we have found that the AMOC recovery can be triggered by the WAIS meltwater discharge, which was already found in a model of similar complexity and given as potential explanation of the Meltwater pulse 1A event leading to the termination of the last glacial period \cite{weaver_meltwater_2003}. Finally, another study showed that the stable AMOC weak state found in the stabilization process corresponds to stadial-like conditions during Dansgaard–Oscher events \cite{willeit_surface_2024}. However, whereas this weak state was found via a variation of ice sheet configuration, greenhouse gas concentration and northern freshwhater flux, our results show that it can be obtained and explained solely via combined ice sheet tipping events.

There are limitations inherent to the simplifications of our study, which require the quantitative results to be interpreted with care. Indeed, the simplified representation of GIS and WAIS meltwater fluxes tipping trajectories, as well as their distribution in space, were not designed to accurately represent future ice sheet tipping trajectories. Yet, these trajectories capture events taking place in the region of interest, and allow to simply and systematically represent the plausible meltwater forcing rates and magnitudes involved in ice sheet tipping events. Also, our experiments did not include interactions from the ocean to the ice sheets, such that impacts of an AMOC collapse on the GIS and WAIS are not present. In fact, the subsequent Northern Hemisphere cooling and Southern Hemisphere warming \cite{jackson_global_2015,orihuela-pinto_interbasin_2022} would likely result in an inhibition and acceleration of the GIS and WAIS meltwater fluxes, respectively \cite{wunderling_climate_2024}. Whereas it is clear that a less brutal GIS meltwater flux may render an AMOC tipping less likely, a faster and/or earlier WAIS tipping may, as we have seen, impact the AMOC resilience. Finally, the critical value of GIS freshwater flux leading to an AMOC collapse (0.035~Sv in the Ghos experiment) is relatively low compared to other studies, for which lower bounds are rather given at about 0.1~Sv. For instance, a study using a more complex model in a comparable experimental setup suggested a critical range of $0.06$~-~$0.16$~Sv \cite{li_global_2023}. Therefore, our experiments should not be directly interpreted as evidence that GIS tipping alone would be sufficient to induce an AMOC collapse.

Our results clearly demonstrate that the AMOC stabilization driven by WAIS meltwater fluxes is not only present in conceptual models, but can also be found in comprehensive climate models. Although the exploratory nature of our study did not permit the use of high-resolution models such as those in the sixth phase of the Coupled Model Intercomparison Project (CMIP6), the use of the well-established EMIC CLIMBER-X provides confidence in the representation of the simulated processes. This is particularly true given the fact that the  climate response to meltwater fluxes in both hemispheres aligns well with other model studies, also involving climate models of higher complexity. This is encouraging for further studies using models of both similar and higher complexity, which are still needed to quantify uncertainties and provide insights into the robustness of the mechanisms involved in the stabilization process. 

The crucial impact of melting rates on AMOC resilience highlighted in our study underscores the necessity of producing long-term projections of ice sheet evolution under diverse emission scenarios, as well as developing advanced coupled models that integrate ice sheet dynamics, oceanic processes, and their interactions. Additionally, reliable observations are crucial not only to monitor the rate of change in meltwater fluxes, but also to diagnose the AMOC state and, given recent advances in early warning signals \cite{van_westen_physics-based_2024}, proximity to tipping.

Our results show that diverse influences of WAIS meltwater on the AMOC can coexist in a single model. Hence, while we emphasise on the beneficial role that a WAIS tipping can have, such a dramatic event is far too dangerous to bet on. Hence, it does not undermine the need for mitigation efforts necessary to avoid any tipping event in the first place. Nonetheless, the profound implications of a prevented AMOC tipping driven by WAIS meltwater for future climate and necessary adaptation make it essential to consider, and investigate further.

\subsection*{Methods}

\subsection*{The climate model}\label{sec:model}
We use the EMIC Climber-X v1.0, which is extensively described in \cite{willeit_earth_2022}. It includes the semi-empirical statistical-dynamical atmosphere model SESAM, the 3D frictional-geostrophic balance ocean model GOLDSTEIN, the sea ice model SISIM and the land model PALADYN, which are all discretized on a $5~^\circ\times 5~^\circ$ horizontal grid and are systematically ran on a yearly time resolution. Each simulation branches off from the end of a pre-industrial control run at which the AMOC is in a monostable regime.

\subsection*{The meltwater forcing}\label{sec:forcing}
An idealized representation of collapsing ice sheets is introduced, analogous to what was done in \cite{sinet_amoc_2023-1}. Namely, we use meltwater flux trajectories that are parabolic in time, allowing to conceptually capture the total duration of each ice sheet tipping event, as well as their delay in time. These trajectories are noted $F_\mathrm{GIS,WAIS}(t)$ for the GIS and WAIS meltwater flux, respectively, and are given by
\begin{equation}
F_{\mathrm{GIS}}(t;P_{\mathrm{GIS}}) = -\frac{6V_{\mathrm{GIS}}}{P_{\mathrm{GIS}}^3}t(t-P_{\mathrm{GIS}})
    \label{eqforcing1}
\end{equation}
if $0<t<P_{\mathrm{GIS}}$ and zero otherwise, and
\begin{equation}
F_{\mathrm{WAIS}}(t;P_{\mathrm{WAIS}},\Delta t) = -\frac{6V_{\mathrm{WAIS}}}{P_{\mathrm{WAIS}}^3}(t-\Delta t)(t-\Delta t - P_{\mathrm{WAIS}})
    \label{eqforcing2}
\end{equation}  
if $0<t-\Delta t<P_{\mathrm{WAIS}}$ and zero otherwise. There, $t$ is the time in years, $P_\text{GIS,WAIS}$ is the duration of each ice sheet tipping event, while $V_\text{GIS,WAIS}$ is the freshwater content of the GIS and WAIS, as given in \cite{sinet_amoc_2023}. By convention, the GIS collapse is always initiated at year $t=0$. This meltwater forcing is applied as a surface, virtual salinity flux without compensation, capturing a full collapse of both the GIS and WAIS. Meltwater fluxes are uniformly spread among grid cells surrounding the GIS and WAIS, as shown in Figure \ref{fig:separate}.a. These regions have been chosen as follows. In both cases, meltwater flux is applied to the first ocean cells directly above, below, or beside ocean-free cells covering the GIS or WAIS ice sheet regions. In the case of GIS hosing region (Figure \ref{fig:separate}.a, left), such cells are only selected under 80.0~$^\circ$N, informed by the procedure depicted by \cite{gerdes_sensitivity_2006} and used in the framework of the North Atlantic Hosing Model Intercomaprison Project (NAHosMIP) \cite{jackson_understanding_2023}. In the case of the WAIS hosing region (Figure \ref{fig:separate}.a, right), such cells are selected between 167.5~$^\circ$E and 27.5~$^\circ$O, i.e. approximately from the beginning of the Ross ice shelf to the end of the Ronne-Filchner ice shelf. 

\subsection*{The freshwater balance}\label{sec:balance}
The freshwater budget over the NAtl box is defined as the time derivative of the total amount of freshwater it contains, here denoted $W$, and given by
\begin{equation}
	\frac{\mathrm{d}W}{\mathrm{d}t} = \Delta F_\text{adv} + \Delta F_\text{diff} + F_\text{surf}.
\end{equation}    
Here, $F_\text{surf}$ is the total surfacer freshwater flux into the box at its surface including, for example, precipitation and runoff. Meanwhile, $F_\text{adv,diff}$ refer to the total northern freshwater transports within the Atlantic due to advection and diffusion at some given latitude, such that $\Delta F_\text{adv,diff} =F_\text{adv,diff}(45~^\circ\text{N}) - F_\text{adv,diff}(60~^\circ\text{N})$ describes the variation of freshwater content within the NAtl box due to each components. Finally, the overturning and azonal components result from a commonly used decomposition of the advective transport (see e.g. \cite{huisman_indicator_2010}), providing the contribution of both the overturning and gyre circulation to the advective transport.

%%%%%%%%%%%%%%%% MAIN TEXT FIGURES %%%%%%%%%%%%%%%

%%%%%%%%%%%%%%%% MAIN TEXT TABLES %%%%%%%%%%%%%%%

%%%%%%%%%%%%%%%% REFERENCES %%%%%%%%%%%%%%%

\clearpage % Clear all remaining figures and tables then start a new page

% The list of references goes after the main text and before the acknowledgements
% When preparing an initial submission, we recommend you use BibTeX, like this:
%
\bibliography{bibli.bib} % for a file named science_template.bib
\bibliographystyle{sciencemag}

% After the paper has completed peer review and been revised ready for acceptance,
% you should comment out the lines above and copy-paste the contents of your .bbl
% file here instead. This will help ensure that our conversion software works correctly.
% Remember to re-run BibTeX first - check the timestamp!
%
% Example of the first three entries copy-pasted from science_template.bbl:
%
%\begin{thebibliography}{1}
%
%\bibitem{example}
%A.~N. {Author}, An example reference. \emph{Journal of Improbable Research}
%  \textbf{1}, 67 (2020).
%
%\bibitem{example2}
%F.~M. {Surname}, S.~{Author}, A second example. \emph{Interesting Research
%  Letters} \textbf{32}, 897 (2019).
%
%\bibitem{example_preprint}
%P.~{One}, P.~{Two}, P.~{Three}, {An unpublished preprint}. \emph{preprint}
%  (2021), arXiv:2101.12345.
%
%\end{thebibliography}

%%%%%%%%%%%%%%%% ACKNOWLEDGEMENTS %%%%%%%%%%%%%%%

\section*{Acknowledgments}

\paragraph*{Funding:}
S.S., A.S.v.d.H. and H.A.D. have received funding from the European Union's Horizon 2020 research and innovation programme under the Marie Sk\l odowska-Curie Actions (grant no. 956170; CriticalEarth). A. S. von der Heydt acknowledges funding by the Dutch Research Council (NWO) through the NWO-Vici project "Interacting climate tipping elements: When does tipping cause tipping?" (project no. VI.C.202.081).
\paragraph*{Author contributions:}
 S.S., A.S.v.d.H. and H.A.D. conceived the idea
for this study. S.S. performed the simulation with CLIMBER-X, conducted the analyses, prepared the figures and wrote the first manuscript. A.S.v.d.H. and H.A. supervised the study and acquired the funding. All authors were actively involved in the interpretation of the analysis results and the writing process.
\paragraph*{Competing interests:}
There are no competing interests to declare.
\paragraph*{Data and materials availability:}
The source code of CLIMBER-X v1.0 can be found on Zenodo at the address https://doi.org/10.5281/zenodo.7898797. All the data produced and used in this study, along with Julia scripts used to analyse the data and produce plots have been archived on Zenodo at the address https://doi.org/10.5281/zenodo.14800555 \cite{sinet_west_2025}.

%%%%%%%%%%%%%%%% SUPPLEMENT LIST %%%%%%%%%%%%%%%

% List the contents of your Supplementary Materials, including the numbers of any
% supplementary figures, tables, external data files etc. and any references that are
% cited only in the supplement. In this example, refs. 7-8 are cited only in the supplement.
% Fill out your numbers accordingly and delete any lines that aren't applicable.
\subsection*{Supplementary materials}
Figures S1 to S3\\
Table S1\\

%%%%%%%%%%%%%%%% END OF MAIN TEXT %%%%%%%%%%%%%%%

\newpage

%%%%%%%%%%%%%%%% START OF SUPPLEMENT %%%%%%%%%%%%%%%

% Figures, tables, equations and pages in the supplement are numbered S1, S2 etc.
\renewcommand{\thefigure}{S\arabic{figure}}
\renewcommand{\thetable}{S\arabic{table}}
\renewcommand{\theequation}{S\arabic{equation}}
\renewcommand{\thepage}{S\arabic{page}}
\setcounter{figure}{0}
\setcounter{table}{0}
\setcounter{equation}{0}
\setcounter{page}{1} % not 0 as \newpage already started a supplementary page
% References continue the numbering from the main text.

%%%%%%%%%%%%%%%% SUPPLEMENT TITLE PAGE %%%%%%%%%%%%%%%

\begin{center}
\section*{Supplementary Materials for\\ \scititle}

% Author list for the supplement
% Indicate the corresponding authors, but do NOT include institutions here
% It would be nice if the template auto-generated this, but doing so is complicated...
	% You can write out first names or use initials - either way is acceptable, but be consistent
Sacha Sinet$^{1,2\ast}$,
Anna. S. von der Heydt$^{1,2}$,
Henk A. Dijkstra$^{1,2}$\\
\small$^{1}$Department of Physics, Institute for Marine and Atmospheric research, \\ \small Utrecht University,  Utrecht, 3584 CC, the Netherlands.\\
\small$^{2}$Centre for Complex Systems Studies, Utrecht Univeristy, Utrecht, 3584 CC, the Netherlands.\\
\small$^\ast$Corresponding author. Email: s.a.m.sinet@uu.nl 
\end{center}

% Fill out the numbers for each type of supplementary material,
% and delete any lines that aren't applicable.
% These are just example numbers that don't match the rest of this template.
\subsubsection*{This PDF file includes:}
Figures S1 to S3\\
Table S1\\

\newpage

%%%%%%%%%%%%%%%% MATERIALS AND METHODS %%%%%%%%%%%%%%%

%%%%%%%%%%%%%%%% SUPPLEMENTARY TEXT %%%%%%%%%%%%%%%
% If your supplement is very short you might need to uncomment the following line to avoid
% layout problems with the figures and tables.
%\newpage

%%%%%%%%%%%%%%%% SUPPLEMENTARY FIGURES %%%%%%%%%%%%%%%

\begin{figure}
    \centering
    \includegraphics[width=1.0\textwidth]{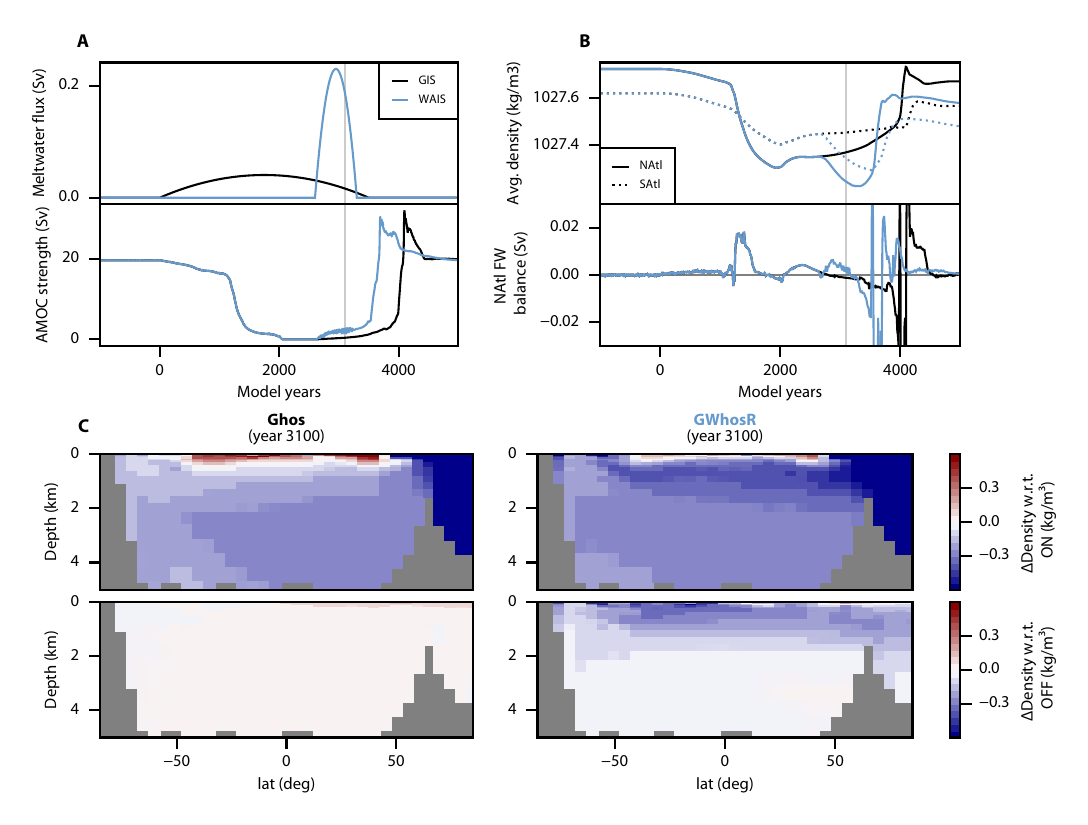}
    \caption{\textbf{AMOC recovery induced by West Antarctic meltwater.} (\textbf{A}) Trajectory of the meltwater fluxes and AMOC strength in the Ghos (black) and GWhosR (blue) experiments, the latter including both the GIS and WAIS meltwater fluxes. (\textbf{B}) Average density of the NAtl and SAtl boxes (above) and freshwater balance of the NAtl box (below), displayed for both the Ghos (black) and GWhosR (green) experiments. (\textbf{C}) Yearly averaged, zonally integrated density anomaly in the Atlantic, with respect to the ON and OFF state (above and below, respectively) for both the Ghos and GWhosR experiments (left and right, respectively). These are displayed at year 3100, depicted by the vertical grey line on A and B.}
    \label{fig:reco}
\end{figure}

\begin{figure}
    \centering
    \includegraphics[width=1.0\textwidth]{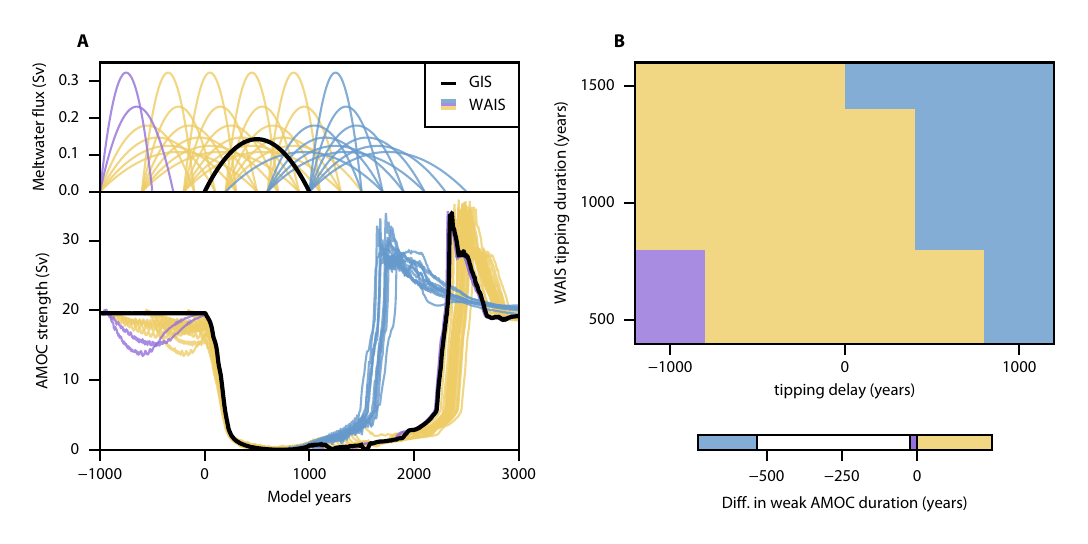}
    
    \caption{\textbf{AMOC response to combined GIS and WAIS meltwater fluxes for a shorter GIS tipping trajectory.} (\textbf{A}) Trajectories of the meltwater fluxes (above) and AMOC (below), fixing the GIS tipping duration to 1000~years. The black trajectory is forced only by the GIS meltwater flux, while these in colours include meltwater from both the GIS and WAIS. Colours represent different values of the difference in weak AMOC duration $\Delta T_\text{weak}$. (\textbf{B}) Difference in weak AMOC duration $\Delta T_\text{weak}$ for trajectories displayed in A, represented in the WAIS tipping durations and tipping delays parameter plane.}
    \label{fig:ressupp}
\end{figure}

\begin{figure}
    \centering
    \includegraphics[width=1.0\textwidth]{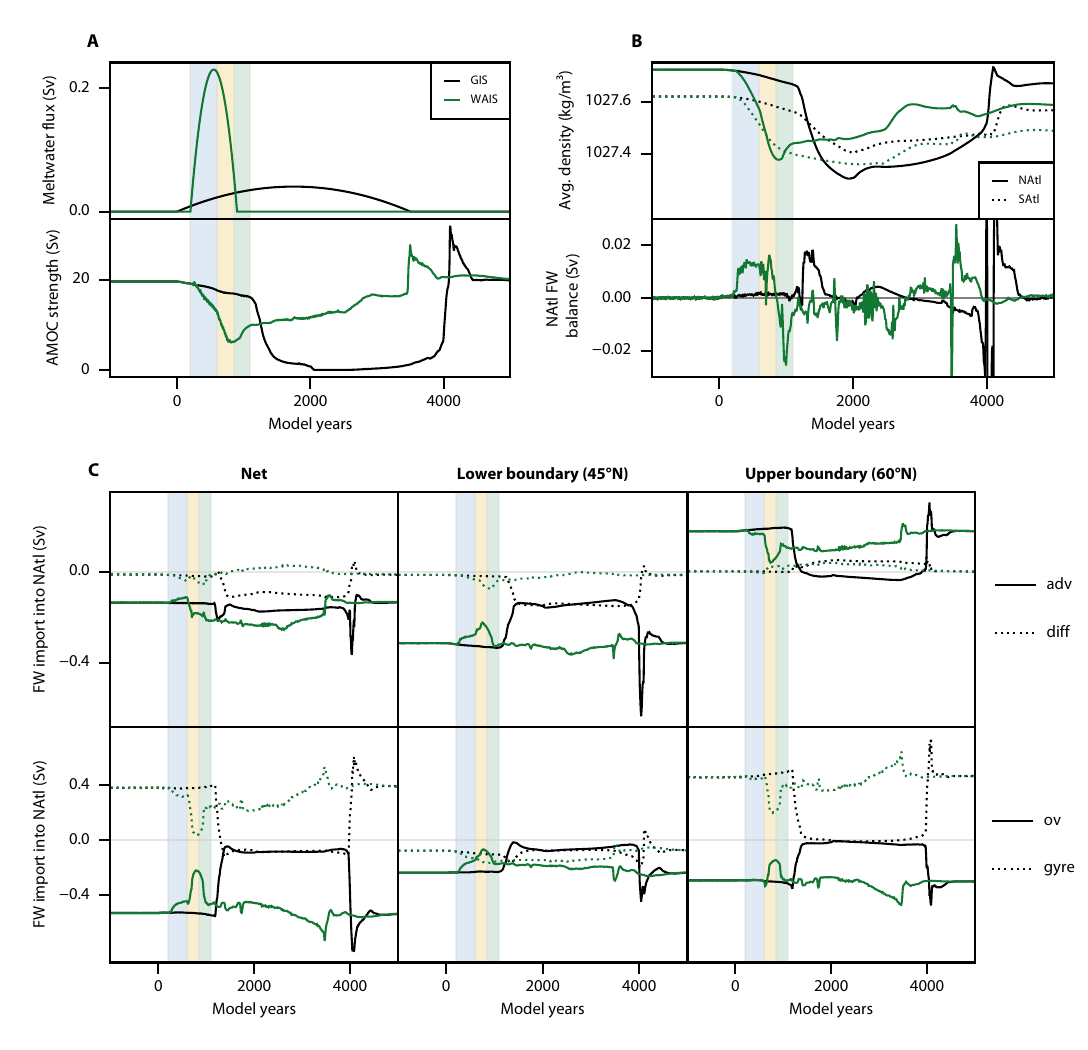}
    
    \caption{\textbf{AMOC stabilization driven by West Antarctic meltwater with more detailed freshwater contributions.} (\textbf{A}) Trajectory of the meltwater fluxes and AMOC strength in the Ghos (black) and GWhosS (green) experiments, the latter including both the GIS and WAIS meltwater fluxes. (\textbf{B}) Average density of the NAtl and SAtl boxes (above) and freshwater balance of the NAtl box (below), displayed for both the Ghos (black) and GWhosS (green) experiments. (\textbf{C}) Net freshwater import into the NAtl box (left), along with the import into the Natl box at its lower (middle) and upper (right) boundaries. It includes the contributions of the advection and diffusion (above) and contributions from both the overturning and gyres (under). These are displayed for both Ghos (black) and GWhosS (green) experiments.}
    \label{fig:FWsupp}
\end{figure}

%%%%%%%%%%%%%%%% SUPPLEMENTARY TABLES %%%%%%%%%%%%%%%
\setcounter{table}{0}
\renewcommand{\tablename}{Table}
\renewcommand{\thetable}{S\arabic{table}}
~
\begin{table}
\centering
\caption{\textbf{Summary of experiments.} Each hosing experiment incorporates ice sheet tipping trajectories with varying tipping durations and, when applicable, different delays between the GIS and WAIS tipping. The Ghos and Whos experiments include only the GIS or WAIS tipping, respectively. The GWhosR and GWhosS experiments involve both the GIS and WAIS tipping, resulting in either AMOC stabilization (GWhosS) or early AMOC recovery (GWhosR).}
\label{tab:exp}

\begin{tabular}{lccc} 
  \\
\hline
  Experiment& GIS tip. duration & WAIS tip. duration & Tip. delay \\ 
  &(years)&(years)&(years)\\
  \hline
  Ghos & 3500 & n.a. & n.a.\\ 
   Whos & n.a. & 700 & n.a.\\ 
  GWhosS & 3500 & 700 & 200 \\ 
  GWhosR & 3500 & 700 & 2600 \\ 
  \hline
\end{tabular}
\end{table}

%%%%%%%%%%% CAPTIONS FOR OTHER SUPPLEMENTARY FILES %%%%%%%%%%

%%%%%%%%%%%%%%%% SUPPLEMENTARY REFERENCES %%%%%%%%%%%%%%%

% Do NOT include a reference list in the supplement.
% All references must be in a single list at the end of the main text.
% The copyeditors will ensure that the correct reference list appears with each version of the paper
% (print, HTML, PDF, mobile app, metadata for bibliographic databases etc.)

\end{document}